\documentclass[journal=jpcbfk, manuscript=article, layout=twocolumn]{achemso}

\usepackage[latin1]{inputenc}
\usepackage{amsmath,amssymb}
\usepackage{graphicx}
\usepackage{siunitx}
\usepackage{xspace}
\newcommand*{\fig}[1]{Fig.~\ref{fig:#1}}
\newcommand{\piprox}{P\textit{i}PrOx\xspace}

\author{Federico Tomazic}
\affiliation{Institute for Multiscale Simulation, Friedrich-Alexander-Universität Erlangen-Nürnberg, 91058 Erlangen, Germany}

\author{Aswathy Muttathukattil}
\affiliation{Institute for Multiscale Simulation, Friedrich-Alexander-Universität Erlangen-Nürnberg, 91058 Erlangen, Germany}

\author{Afshin Nabiyan}
\affiliation{Jena Center for Soft Matter, Friedrich-Schiller-Universität Jena, 07743 Jena, Germany.}
\alsoaffiliation{Institute of Organic Chemistry and Macromolecular Chemistry, Friedrich-Schiller-Universität Jena, 07743 Jena, Germany.}
\alsoaffiliation{Center for Energy and Environmental Chemistry, Friedrich-Schiller-Universität-Jena, 07743 Jena, Germany.}

\author{Felix Schacher}
\affiliation{Jena Center for Soft Matter, Friedrich-Schiller-Universität Jena, 07743 Jena, Germany.}
\alsoaffiliation{Institute of Organic Chemistry and Macromolecular Chemistry, Friedrich-Schiller-Universität Jena, 07743 Jena, Germany.}
\alsoaffiliation{Center for Energy and Environmental Chemistry, Friedrich-Schiller-Universität-Jena, 07743 Jena, Germany.}

\author{Michael Engel}
\email{michael.engel@fau.de}
\affiliation{Institute for Multiscale Simulation, Friedrich-Alexander-Universität Erlangen-Nürnberg, 91058 Erlangen, Germany}

\title{Geometric Frustration Directs the Self-Assembly of Nanoparticles with Crystallized Ligand Bundles}


\begin{document}

\begin{tocentry}
	\includegraphics[width=1\textwidth]{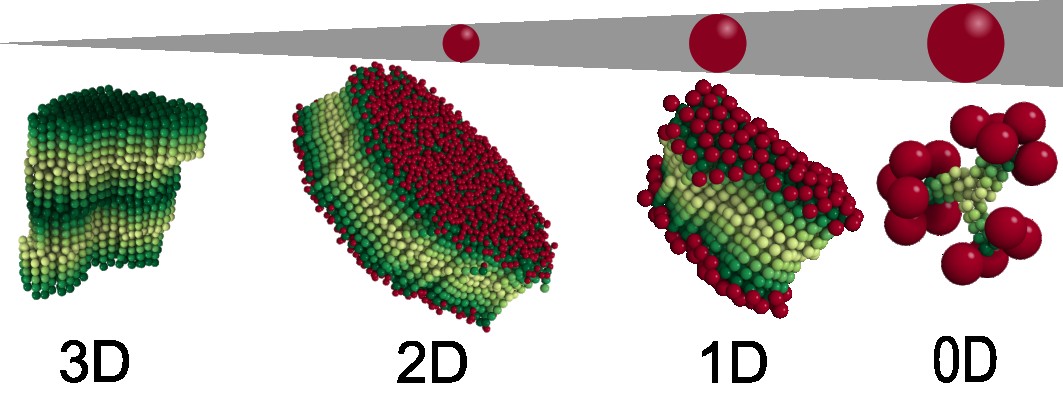}
\end{tocentry}

\begin{abstract}
Polymer-grafted nanoparticles are versatile building blocks that self-assemble into a rich diversity of mesostructures. Coarse-grained molecular simulations have commonly accompanied experiments by resolving structure formation pathways and predicting phase behavior. Past simulations represented nanoparticles as spheres and the ligands as flexible chains of beads, isotropically tethered to the nanoparticles. Here, we investigate a different minimal coarse-grained model. The model consists of an attractive rod tethered to a repulsive sphere. The motivation of this rod-sphere model is to describe nanospheres with a partially crystallized, stretched polymeric bundle, as well as other complex building blocks such as rigid surfactants and end-tethered nanorods. Varying the ratio of sphere size to rod radius stabilizes self-limited clusters and other mesostructures of reduced dimensionality. The complex phase behavior we observe is a consequence of geometric frustration.
\end{abstract}

\section{Introduction}

Polymer-grafted (`hairy') nanoparticles and the mesostructures they self-assemble have many applications, including as electronic and optical materials, in medicine, and biological imaging.\cite{kumar_nanocomposites_2013, yi_polymer-guided_2020, bassani_nanocrystal_2024}
Structure formation with such nanoparticles is usually controlled by tailoring the properties, density, and type of polymer ligands.\cite{chancellor_characterizing_2019, he_engineering_2023}
The underlying processes have been studied in molecular simulations by modeling nanoparticles as rigid spheres and the ligands as flexible bead-chains, uniformly distributed over the nanoparticle surface. \cite{pryamtisyn_modeling_2009, akcora_anisotropic_2009, koh_assembly_2020, borowko_shape_2022, nabiyan_self-assembly_2023} Bead-chain models predicted various types of mesostructures, including fibers, platelets, and three-dimensional agglomerates, depending on the polymer density and the length of the grafted chains. \cite{akcora_anisotropic_2009} When the distribution of chains is not uniform, more structures can be obtained, such as cylinders, perforated lamellae and cubic ordered micelles. \cite{glotzer_self-assembly_2005, iacovella_phase_2005}\cite{moinuddin_effect_2022, cui_solution_2023}
The interaction between ligands can be understood via the optimal packing model,\cite{landman_small_2004} the optimal cone model,\cite{schapotschnikow_understanding_2009} or the orbifold topological model.\cite{travesset_soft_2017, waltmann_many_2018} These models make assumptions about the ligand shell. Specifically, the optical cone model and the orbifold topological model consider many-body ligand interactions, and the orbifold topological model also considers polymer bundling in form of vortices. Similar bundling is frequently observed in flexible bead-chain simulations.\cite{kister_colloidal_2018, santos_dictating_2019}

While flexible bead-chain bundles always remain flexible, some polymers can order more strongly or even crystallize directionally. Poly(2-isopropyl-2-oxazoline) (\piprox) is such an example. \piprox  forms porous spheroidal clusters in water, out of which nanofibers sprout through directional crystallization. \cite{demirel_formation_2007, diehl_mechanistic_2010, oleszko-torbus_thermal_2019}
Polymer backbones align in the ribbons, separating the isopropyl groups on one side from the amide groups on the other side. The side with the isopropyl group grows faster. It was therefore suggested that the anisotropic growth of \piprox  nanoribbons is due to the combination of the hydrophobic unspecific interaction of the isopropyl groups and the dipole interactions of the amine groups.\cite{demirel_formation_2007, sun_globules_2015, oleszko-torbus_thermal_2019} Grafting \piprox  to silica nanospheres is a strategy towards complex self-assembled mesostructures, \cite{willinger_thermoresponsive_2021, nabiyan_self-assembly_2023}
Because \piprox  orders so strongly, the bead-chain model is no longer valid. Predicting the mesostructures, which may be self-assembled with building blocks such as \piprox is the motivation for the present work.

\begin{figure}
    \centering
	\includegraphics[width=1\columnwidth]{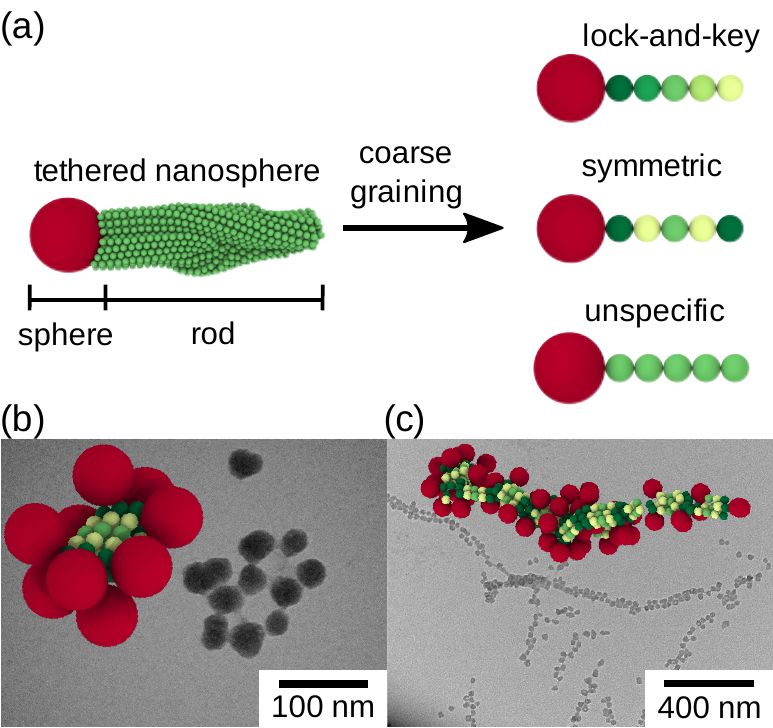}
	\caption{\textbf{Tethered nanoparticle building block modelled by the rod-sphere particles.} (a) Coarse-grained model of a spherical nanoparticle with a partially crystallized grafted polymer bundle and further coarse-graining maps the polymer bundle to a single rod in form of a rigid bead chain. We distinguish three rod types depending on the interaction of the individual beads as represented by the shades of green color. The types are (from top to bottom), (i)~\textit{lock-and-key} -- all beads are different, (ii)~\textit{symmetric} -- the sequence of beads is a palindrome, (iii)~\textit{unspecific} -- all beads are identical. (b) and (c) TEM micrographs showing the self-assembly of \piprox-grafted silica nanoparticles forming small aggregates (b) and fibers (c). Similar structures obtained in the simulations using the rod-sphere model are superimposed.}
	\label{fig:model}
\end{figure}

We model the silica nanoparticle with its homogeneously grafted non-crystalline polymer shell as a sphere and a stretched crystalline polymer bundle as a rigid rod. We then tether the rod to the sphere to obtain a coarse-grained model particle that resembles a surfactant (\fig{model}a). We call this particle model a rod-sphere particle. With this model, it is possible to reproduce the structures that \piprox-grafted silica nanoparticles in aqueous solution self-assemble in experiment. Transmission electron microscopy (TEM) micrographs of small aggregates and fibers and corresponding simulation snapshots are shown in \fig{model}b and \fig{model}c, respectively. The fiber-like structures are obtained by heating the solution for 24h to 65$^{\circ}$C.\cite{nabiyan_self-assembly_2023}

 Surfactant self-assembly into micelles of various shapes has been described by considering the aspect ratio between the hydrophilic head and the hydrophobic tail. With increasing ratio of the size of the tail to the size of the head, and at constant interaction with the solvent, the mesostructures transition from spherical micelles to cylindrical micelles to lamellae.\cite{nisraelachvili_theory_1976, nagarajan_molecular_2002} The simulations of the present work focus on a similar geometric effect. As we will show in our model system, the sphere head induces geometric hindrance that frustrates the formation of bulk three-dimensional structures. Geometrical frustration then causes self-limiting assembly into mesostructures with finite spatial extent.\cite{lenz_geometrical_2017, hagan_equilibrium_2021}

Besides nanospheres with crystallized ligand bundles, the rod-sphere model can describe a number of other experimentally-relevant nanoscale building blocks. First, the rod-sphere model can reproduce the self-assembly of surfactants, including giant surfactants\cite{yu_giant_2010} and block copolymers, which have a rigid lyophobic segment. Amphiphilic particles with a rigid tail have applications as stabilizer agents for single wall carbon nanotube dispersions \cite{dutta_ph-responsive_2012, huang_light-controlled_2012, yan_co2-responsive_2018} or as photoresponsive foams.\cite{lei_photoresponsive_2017}In particular, \piprox can be used as one block of the copolymer.\cite{kim_dual_2012, jana_poly2-oxazoline-based_2020} Second, the rod-sphere model can mimic rod-shaped nanoparticles tethered unilaterally with polymers.\cite{petukhova_standing_2012} The model is then inverted, i.e.\ the rod represents the nanoparticle and the sphere the aggregated polymers. Similar simulations representing polymers as chains of beads showed the organization of particles in mesostructures of various shapes, depending on the aspect ratio between the rod and the chain.\cite{horsch_self-assembly_2005,horsch_self-assembly_2006, horsch_simulation_2006, he_solvent-induced_2010, horsch_self-assembly_2010} 

\section{Model and Methods}

\subsection{Rod-sphere model}
A single tethered nanosphere is modeled as a rod-sphere particle with two main components (\fig{model}a), a sphere of radius $R$ representing the nanosphere (red) and a rigid rod representing the ligands bundled tether (green). The rod then represents the part of the polymer shell that aggregates and crystallizes. The rigid rod is a short chain of five spherical beads of radius $r$, which can translate and rotate but not bend or stretch. Each bead is assigned a bead type. Beads of the same type interact attractively with a Lennard-Jones potential. Beads of different type interact repulsively with a Weeks-Chandler-Andersen\cite{weeks_role_1971} (WCA) potential. The model is similar to previous models of tadpole particles, nanoparticles where a single polymer chain is attached to a spherical nanoparticle.\cite{lee_self-assembly_2004, moinuddin_effect_2022, cui_solution_2023} The main difference is that in our model is replaced by a rigid rod. The rod is tethered on one end to the sphere using a harmonic bond with a rest length $r_0 = r + R$ and a stiffness coefficient $k=10\varepsilon r^{-2}$, since the polymer ends attached to the particle are spaced sufficiently apart, preventing crystallization in that region and allowing for flexibility. We opted not to include bending interactions in the model for simplicity. Spheres interact repulsively with other spheres and with beads via a WCA potential.

We distinguish three rod types by the choice of the bead types and their interaction: 
(1)~In \textit{lock-and-key} rods, each of the five beads is of different type. The sequence of five beads corresponds to A-B-C-D-E. Different bead types are indicated by different shades of green in \fig{model}a. Rods attract strongly if they are parallel with the same orientation.
(2)~In \textit{unspecific} rods, all beads are identical. The sequence of beads is A-A-A-A-A. Rod-rod attraction between neighboring rods increases gradually as the contact between them increases.
(3)~In \textit{symmetric} rods, the beads form a palindromic sequence, i.e. A-B-C-B-A. Strong rod-rod attraction is achieved when the rods are parallel in either the same or in the opposite orientation. The three rod types represent different interaction modes between polymer chains when they crystallize. The unspecific rod describes the effect of homopolymers, the symmetric and lock-and-key cases capture more complex polymers, such as DNA ligands forming scaffolds.

\subsection{Molecular dynamics simulations}
We simulated rod-sphere particles with molecular dynamics using the \textit{HOOMD-blue} general-purpose particle simulation toolkit (version 2.9.0) on a GPU,\cite{anderson_general_2008, nguyen_rigid_2011, glaser_strong_2015} analyzed configurations with the \textit{freud} Python library\cite{ramasubramani_freud_2020}, and visualized them with the \textit{Ovito} software\cite{stukowski_visualization_2009}.
The Lennard-Jones attraction between beads has the parameters $\varepsilon$ and $\sigma = 2r$. It is truncated at $r_\text{cut}=2.5\sigma$ and shifted to zero. Likewise, the repulsive WCA potential is a shifted Lennard-Jones potential with the same $\varepsilon$, $\sigma$ equal to the sum of two radii, and the cut-off $r_\text{cut}=2^\frac{1}{6}\sigma$.
Interaction of the particles with the solvent was considered implicitly via a Langevin integrator.\cite{bussi_accurate_2007} A cell list was used for most simulations, except when simulating the twist of the lock-and-key ribbons, in which case the linear bounding volume hierarchies tree method was used.\cite{howard_efficient_2016, howard_quantized_2019} Dimensionless units were chosen by setting $r$ and $\varepsilon$ as the units for length and energy. The mass unit $m$ was fixed by setting the density of the beads and the spheres equal to $m/r^3$. The simulation runtime is $t^* = \sqrt{mr^2/\varepsilon}$.

Each simulated system consists of 10648 rod-sphere particles in a cubic box with periodic boundary conditions. The volume fraction was constant at $\phi=0.075$, assuming the radius of the sphere is $R$ and the radius of the rod beads is $r$.
This value is a compromise between typical experimental conditions, where the volume fraction is even lower,\cite{nabiyan_self-assembly_2023}
and keeping concentration high enough to assemble efficiently. Simulations were continued for a runtime of $75000 t^*$ with time step $\delta t = 0.01\sqrt{\varepsilon/kT}t^*$. We initialize the particle positions in a cubic lattice and run a short isothermal-isobaric  simulation at constant pressure to adjust the volume size and randomize the system.

We performed hundreds of simulations, each with a different variation of the rod sphere model. First of all, we used the three different interaction types: lock-and-key, symmetric and unspecific. Secondly, we tested the effect of different nanosphere sizes, using 16 different models with nanosphere radii $R$ ranging from a minimum of $R=0.25r$ to a maximum of $R=4r$. This model correspond either to larger nanoparticles tethered to the same bundle, or to the same nanoparticle tethered to longer polymer bundles. In addition, we performed simulations without the nanosphere, where only untethered rods are present.
Lastly, we ran simulations at 10 different temperatures, ranging from $kT = 0.1\varepsilon$ to $kT = \varepsilon$ for the symmetric and lock-and-key interactions and from $kT = 0.2\varepsilon$ to $kT = 2\varepsilon$ for the unspecific interactions. We keep the temperature in each separate simulation run constant.

We estimate the total simulation runtime to be only about  $5\si{\milli\second}$ by comparing the self-diffusion coefficient in the simulation with the one of a \piprox-grafted silica nanospheres experiment, which have a typical rod length $30\si{\nano\meter}$ and whose diffusion coefficient can be calculated by the Stokes-Einstein equation to be $D\approx4\cdot10^{-11}\si{\square\meter\per\second}$,\cite{edward_molecular_1970}
The achievable simulation runtime is much shorter than the runtime of the experiment. However, the experiment has many more degrees of freedom we do not consider. Polymers must assemble into crystalline bundles before self-assembly can occur while in simulations polymers are already bundles. In addition, the volume fraction chosen in simulation is higher, speeding up the self-assembly process.

\subsection{Mesostructure classification}
\begin{figure*}
    \centering
    \includegraphics[width=0.8\textwidth]{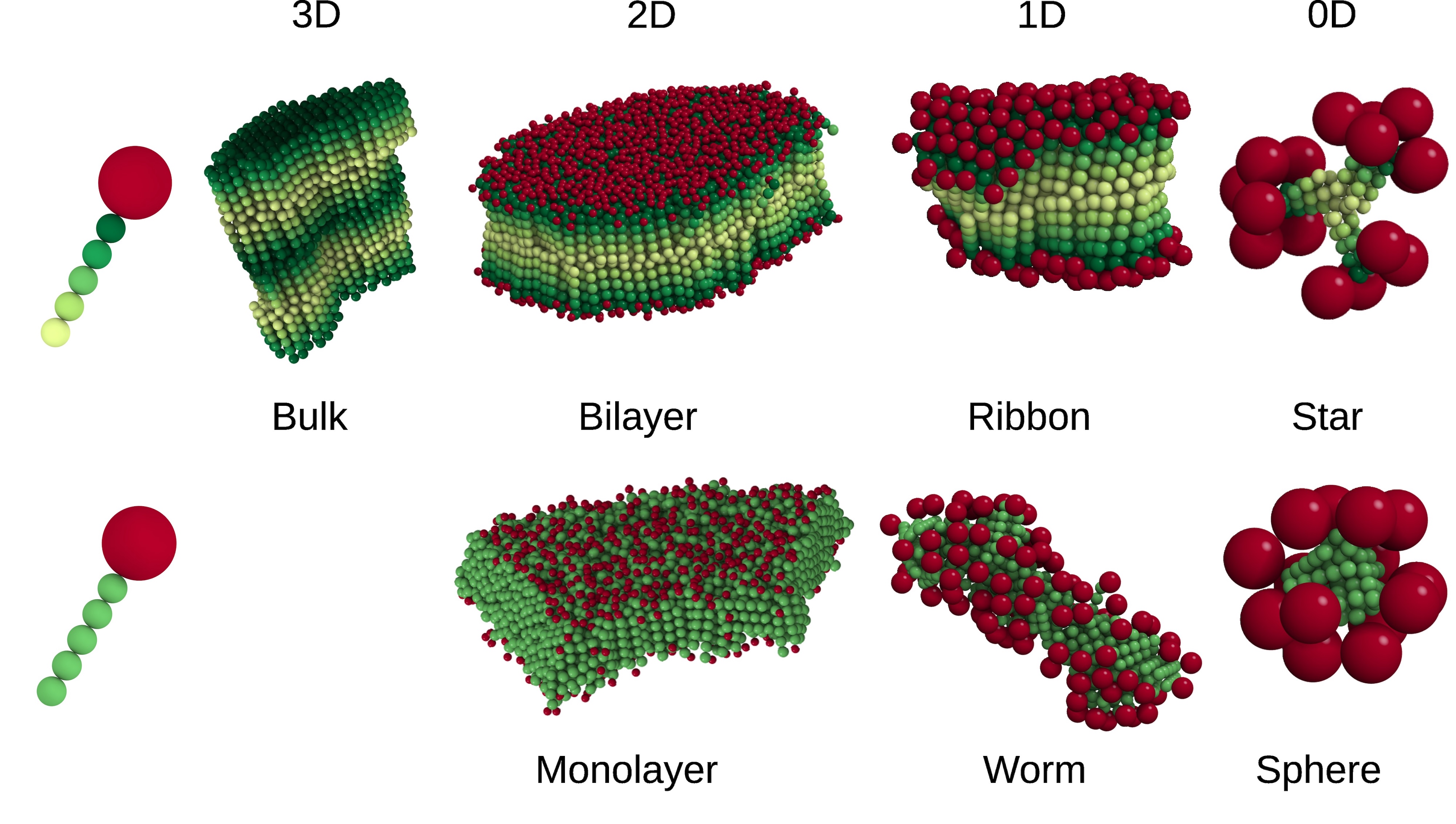}
    \caption{\textbf{Classes of mesostructures of rod-sphere particles.}
    Mesostructures of rod-sphere particles with lock-and-key rods (top row) and with unspecific rods (bottom row). Mesostructures are grouped according to their dimensionalities. Rod-sphere particles with symmetric rods can form mesostructures from both rows.}
    \label{fig:clusters}
\end{figure*}

A number of techniques exist for structure analysis based on local order parameters.\cite{keys_characterizing_2011_1, keys_characterizing_2011_2, duboue-dijon_characterization_2015, lazar_topological_2015}
However, these are not applicable here, because local order in our simulations is often incomplete, similar to liquid crystalline order (\fig{clusters}). Instead, we perform classification in two steps: first, we partition the set of all rod-sphere particles into clusters and second, we determine the dimensionality of the clusters. We consider only the rods in our analysis and neglect the spheres. The geometry of the $i$-th rod is fully specified by its centroid and the direction vector $\hat{d}_i$, which is normalized and points along the rod axes away from the sphere the rod is tethered to.

In a first step (partitioning into clusters), we separate all rods into ordered and disordered rods. A rod is ordered if its neighbors are well aligned with it. Specifically, rod $i$ is said to be ordered, if it fulfills the condition $\frac{1}{N_i}\sum_{j} |\hat{d}_i\cdot\hat{d}_j| \ge 0.99$, where the sum runs over all nearest neighbors. Two rods are nearest neighbors if any of their constitutive beads are closer than $r_0=2.25r$. The ordered and disordered rods are further separated into clusters. Clusters are the connected component of the network obtained by connecting rods if any of their beads are closer than $r_0$.

In a second step (dimensionality analysis), we define the dimensionality of a cluster based on its gyration tensor. The gyration tensor for a cluster $c$ with $N_c$ rod-sphere particles is\cite{theodorou_shape_1985,arkin_gyration_2013}
\begin{equation*}
    S_c=\frac{1}{N_c}
    \begin{bmatrix}
        \sum_ix_i^2 & \sum_ix_iy_i & \sum_ix_iz_i \\ 
        \sum_ix_iy_i & \sum_iy_i^2 & \sum_iy_iz_i \\ 
        \sum_ix_iz_i & \sum_iy_iz_i & \sum_iz_i^2
    \end{bmatrix}.
\end{equation*}
where the sums run over all rods $i\in c$ and $(x_i, y_i, z_i)$ are the rod centroids in a frame of reference that has its origin in the center of mass of the cluster. The gyration tensor has three eigenvalues, $\lambda_{c,1}$, $\lambda_{c,2}$, $\lambda_{c,3}$. An eigenvalue is small if the cluster has narrow extension in the direction of the eigenvector and large if the cluster has a large extension. We define the dimensionality of the cluster as
\begin{equation}
    D_c = 3-\sum_{k=1}^3 \exp\left(-\frac{\lambda_{c,k}}{a}\right),
\end{equation}
In this equation, the exponential tends to 0 if the eigenvalues are large, and tend to 1 if the eigenvalues are small. Therefore, if the structure extends in all directions, all eigenvalues are large and $D\approx3$; if the structure is flat, one eigenvalue is small and $D\approx2$; if the structure is elongated, only one eigenvalue is large and $D\approx1$, and finally if the cluster is small, all the eigenvalues are small and $D\approx0$. The parameter $a$, therefore, specifies a minimum size for an eigenvector (and a structure) to be considered small. Because the average distance between rod centroids is different in a cluster of ordered rods than in a cluster of disordered rods, we choose two values for the minimum size, $a=30r^2$ for ordered rods and $a=130r^2$ for disordered rods. We select the specific value by comparing the $D$ parameter as a function of $a$ with visual observation of the clusters. The parameter $D$ provides a more intuitive and tangible approach to classify structures of different dimensionalities compared to, e.g., asphericity.

We analyze the geometry of the mesostructure in a simulation snapshot using two global parameter: (1)~The \emph{degree of order parameter}, $0\leq O\leq 1$, is the fraction of ordered rod-sphere particles in the simulation snapshot. (2)~The \emph{dimensionality parameter}, $0\leq D\leq 3$, is the weighted average of the dimensionality of all clusters,
\begin{equation}
    D=\frac{\sum_{c\in C} N_c D_c}{\sum_{c\in C} N_c}.
\end{equation}
The parameters $D$ and $O$ extract complementary information from a mesostructure. $D$ quantifies the geometry of clusters, and $O$ quantifies how parallel rods are.

\section{Results and Discussion}
 Rod-sphere particles self-assemble into different classes of mesostructures, depending on the temperature $T$ at which the simulation is performed and on the frustration induced by the spherical parts of the particles, quantified by the geometric parameter size ratio $R/r$. The seven classes of mesostructures observed in this work are summarized in \fig{clusters}. In bulk assembly, particles extend in all directions from the center of mass of the cluster. Therefore the three eigenvalues of the gyration tensor are large and $D\lessapprox 3$. In monolayer and bilayer assemblies, two eigenvalues are much larger than the third, thus $D\approx 2$. Monolayers have a thickness of just one rod, while bilayers have a thickness of two rods. The particle heads appear on both sides. In ribbons, fibers, and worms, only one eigenvalue is larger than the other two, thus $D\approx 1$. In ribbons, rods are parallel and their direction vectors perpendicular to the large eigenvalue. In fibers, rod direction vectors are parallel to the large eigenvalue. In worms, the rod direction vectors are disordered. Finally, in star and sphere clusters, all eigenvalues remain small, $D\gtrapprox 0$. Whereas spherical clusters are rather compact with rods overlapping in various directions, stars are more open with non-parallel rods in contact only at their ends. In addition, we consider three other classes of mesostructures: percolating gel (Fig.~S14), disperse (Fig.~S13), and liquid phase (Fig.~S12). In these mesostructures, particles remain disordered.

\subsection{Evolution of structural order}
\begin{figure*}
    \centering
    \includegraphics[width=0.99\textwidth]{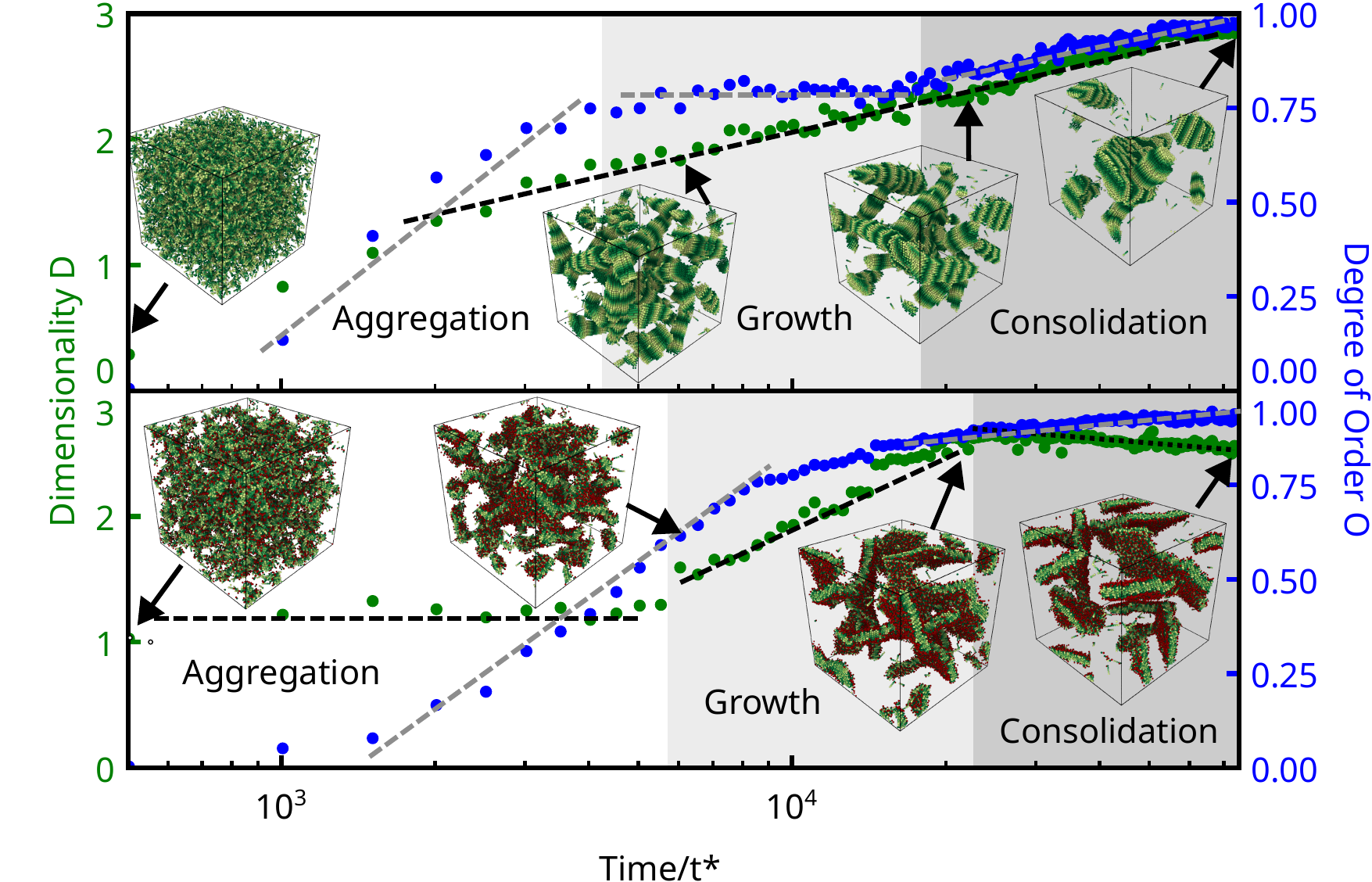}
    \caption{\textbf{Mesostructure order evolution.} Dimensionality parameter $D$ (left side) and degree of order parameter $O$ (right side) along two simulation trajectories. The plots shot the self-assembly of two mesostructures: top:~crystalline bulk (lock-and-key interactions, $R=0$, $kT=0.8\varepsilon$), and bottom: monolayers (unspecific interactions, $R=r, kT=1.2\varepsilon$). Characteristics snapshots are also shown. The cluster formation can be subdivided into three steps: aggregation, growth and consolidation.}
	\label{fig:evolution}
\end{figure*}

To further test the dimensionality parameter $D$ and the degree of order parameter $O$, we use them to analyze two simulation trajectories where crystalline bulk and monolayer structures are formed (\fig{evolution}).  Four snapshots for each simulation are available in the supporting information (SI Figs.~1,2). In both simulations, the formation of ordered mesostructures can be subdivided in three steps. A rapid initial aggregation step is followed by a growth step and a much slower consolidation step. In \fig{evolution} top, the interaction is of lock-and-key type, and we consider simple rods ($R=0$). Fast nucleation from an initial state with $D=O=0$ is followed by gradual growth into a fully ordered three-dimensional cluster approaching $D=3, O=1$. At first, most particles aggregate into thin cylindrical clusters. In the growth phase, the remaining particles attach to the cylinder surface. As clusters become larger and grow in all directions, $D$ increases, while $O$ remains almost constant. Finally, in the consolidation step, cylinders merge together into wider structures, growing laterally. The merging coincides with a decrease in potential energy (Fig.~S15). The order parameter $O$ starts increasing again, since the cluster surface to volume ratio decreases and rods in the cluster bulk are more ordered than rods on the surface. The final three-dimensional clusters are still elongated, but they are not considered one-dimensional since they extend in all directions.

In \fig{evolution} bottom, rod-spheres with unspecific interactions and a sphere of radius $R=r$ form single layer sheets. The presence of the spheres causes the particles to aggregate first in disordered, one-dimensional, worm-like structures with $D=1$. In the aggregation step, the rods inside these thin structures start to orientate in the same direction, increasing $O$ and eventually forming small sheets; In the growth phase, the sheets merge into larger monolayers, and they also grow by the attachment of the remaining disperse rods. In the consolidation state, $D$ decreases slightly, since the clusters extend more and more in two dimensions compared to the third one. One large monolayer forms, comprising most of the particles and wrapping multiple time over the simulation box. The degree of order parameter $O$ increases to $~1$.

In both simulations analyzed, the parameters $O$ and $D$ do not fluctuate during the simulation, even if clusters move in the simulation box, continuously colliding and separating. This analysis demonstrates the robustness of the parameters $D$ and $O$ to classify the dominant mesostructure along the simulation trajectory, identifying the different steps in self-assembly. In addition, the trend of the parameters value helps to distinguish the three steps of aggregation, growth and consolidation. Aggregation is characterized by an increase in the values of $O$, growth by an increase of $D$, and consolidation by a slower increase of $O$.

\subsection{Mesostructure maps}

\begin{figure*}
	\centering
	\includegraphics[width=0.9\textwidth]{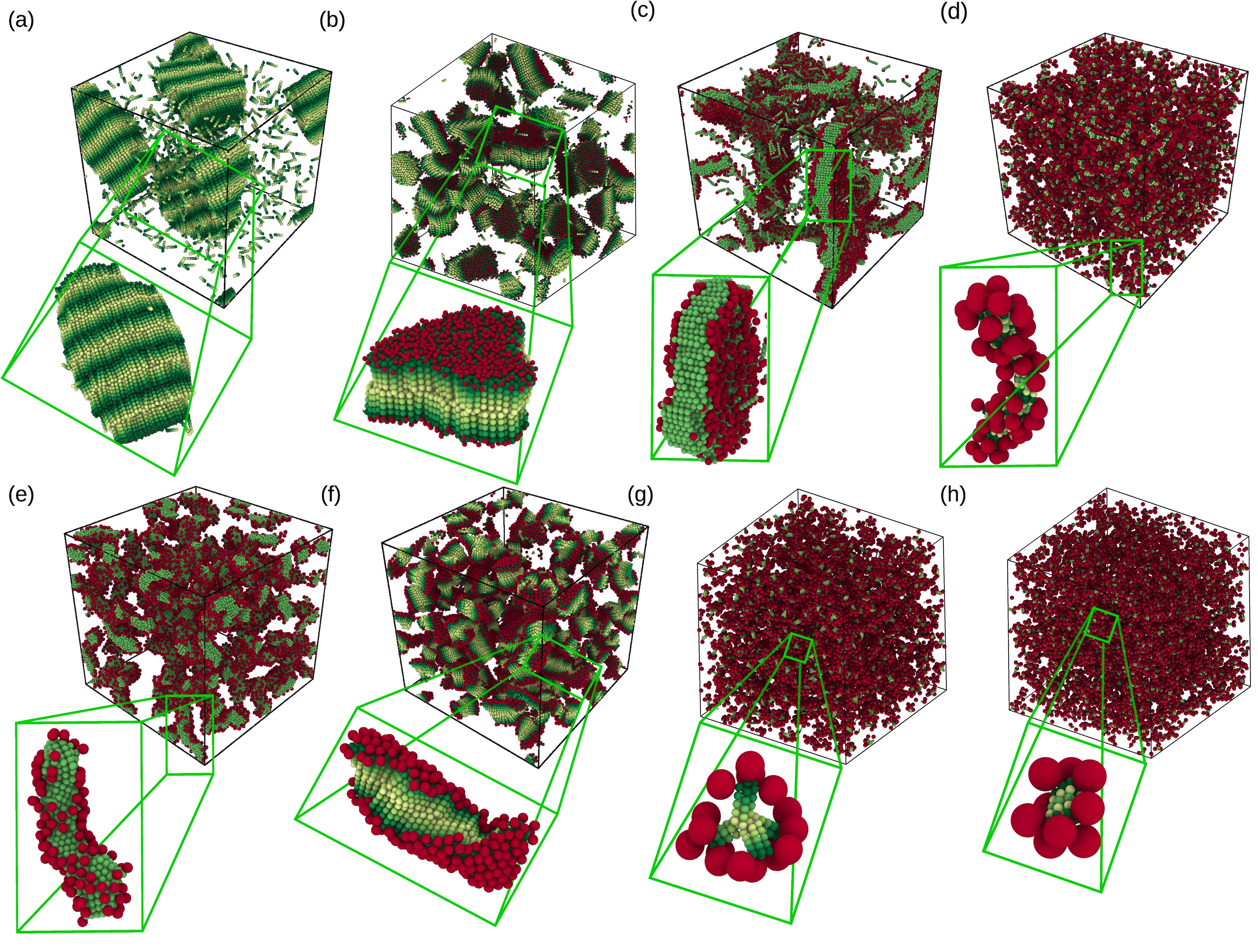}
    \caption{\textbf{Diversity of mesostructures self-assembled form the rod-sphere model.} Snapshots show the last frame of long simulation trajectories. They are sorted in order of decreasing dimensionality. We specify simulation parameters in Table~\ref{tbl:snapshots}. (a) Bulk assembly (3D); (b) Bilayers (2D); (c) Monolayers (2D); (d) Fibers (1D); (e) Worms (1D); (f) Ribbons (1D); (g) Stars (0D); (h) Spheres (1D).}
    \label{fig:snapshots}
\end{figure*}

The self-assembly behavior of rod-sphere particles is dominated by the thermodynamic parameter temperature $T$ and the geometric parameter size ratio $R/r$. To systematically understand mesostructure self-assembly as a function of these parameters, we performed 170
simulations for the three rod types `lock-and-key', `symmetric', and `unspecific'. Exemplary simulation snapshots from each mesostructure type are shown in \fig{snapshots}. Additional simulation snapshots of all mesostructures are available in Supporting Information (Figs.~S3-S14).

As is apparent from these snapshots, the geometric shape of the rod-sphere particle is critical to control the type of the self-assembled mesostructure. The larger the size ratio $R/r$, i.e., the larger the red spheres in \fig{snapshots}, the lower the dimensionality. This behavior agrees well with a similar sequence of structures found with amphiphilic molecules.\cite{nisraelachvili_theory_1976, nagarajan_molecular_2002}

\begin{table*}
  \caption{Dimensionality and order parameters of snapshots in \fig{snapshots}}
  \label{tbl:snapshots}
  \begin{tabular}{llllllll}
    \hline
    snapshot & type & dimensions & D & O & interaction & R / r  & kT / $\varepsilon$ \\
    \hline
    a & Bulk assembly & 3D & 2.82 & 0.93 & lock-and-key & 0.25 & 0.9 \\
    b & Bilayers      & 2D & 2.06 & 0.92 & lock-and-key & 0.75 & 0.7 \\
    c & Monolayers    & 2D & 2.08 & 0.66 & unspecific   & 1.25 & 1.2 \\
    d & Fibers        & 1D & 1.13 & 0.17 & symmetric    & 2.5  & 0.1 \\
    e & Worms         & 1D & 1.22 & 0.11 & unspecific   & 1.5  & 0.8 \\
    f & Ribbons       & 1D & 1.10 & 0.65 & lock-and-key & 1.25 & 0.4 \\
    g & Stars         & 0D & 0.22 & 0.03 & lock-and-key & 3.5  & 0.4 \\
    h & Spheres       & 0D & 0.18 & 0.54 & symmetric    & 3.5  & 0.3 \\
    \hline
  \end{tabular}
\end{table*}

\begin{figure*}
    \centering
	\includegraphics[width=\textwidth]{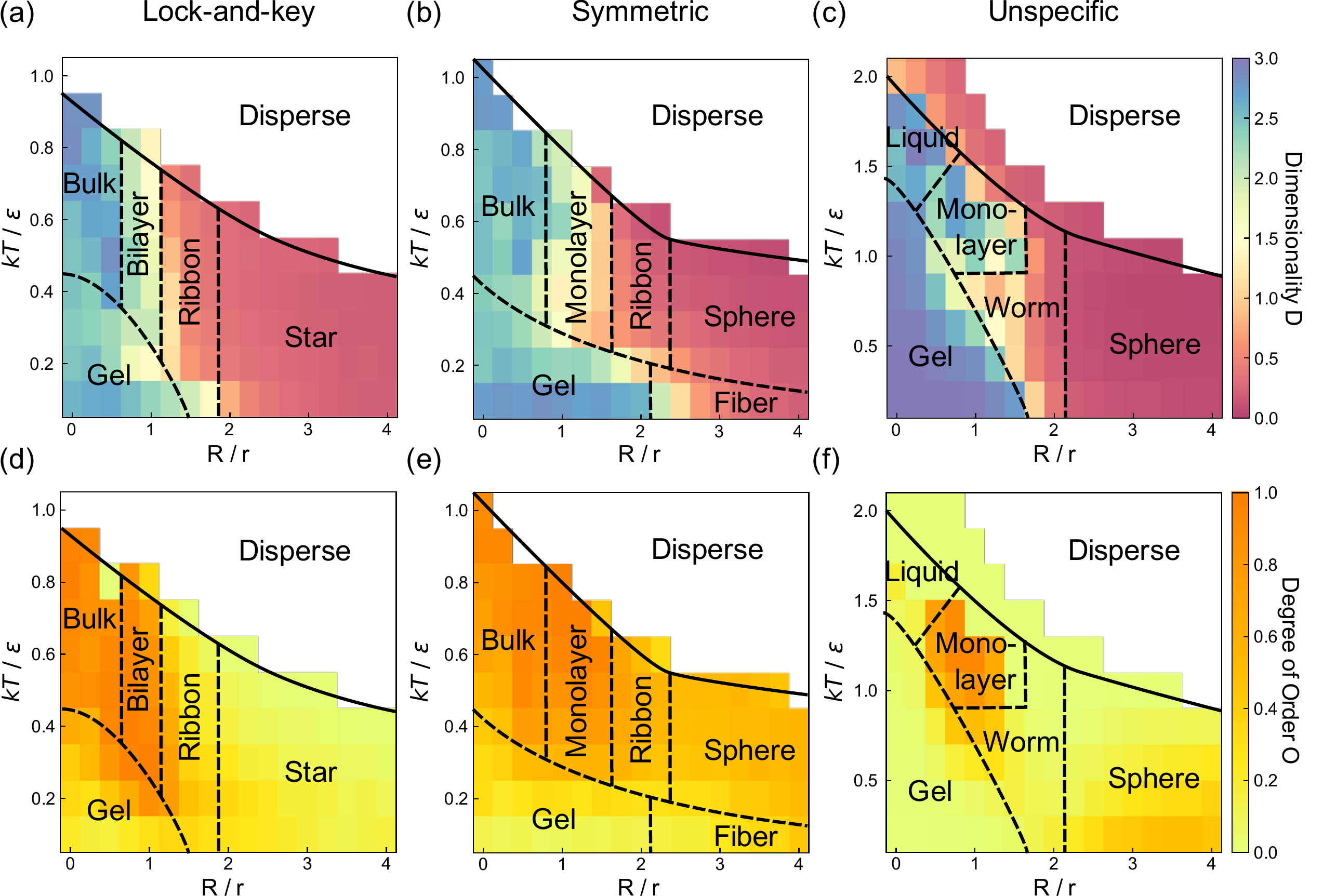}
    \caption{\textbf{Mesostructure maps quantifying dimensionality and the degree of order.} (a-c) The dimensionality parameter $D$ decreases gradually with increasing the nanosphere size. (d-f) By combining it with the degree of order parameter $O$, the dominant mesostructure in each simulation can be uniquely deduced. (a,d) lock-and-key rods, (b,e) symmetric rods, (c,f) unspecific rods. Lines are guides to the eye.}
	\label{fig:maps}
\end{figure*}

We build mesostructure maps for the three rod types by analyzing all final simulation snapshots with the dimensionality parameter $D$ and the degree of order parameter $O$.  The results are shown in \fig{maps}. As expected, dimensionality maps (\fig{maps}a-c) confirm that $D$ generally decreases as the sphere size $R$ increases. We can thus separate mesostructures into four classes by their dimensionality (four columns in \fig{clusters}). To complete the classification and distinguish mesostructures with the same dimensionality, we resort to the degree of order parameter maps (\fig{maps}d-f). We complement the parameters by visually classifying the final structures (black lines). The gradual decrease of $D$ with larger sphere size $R$ shows that there is not a clear-cut transition between different structure types. We can still identify different structures using $D$: if $2.5<D<3$, three-dimensional; if $1.5<D<2.5$, two-dimensional; if $0.5<D<1.5$, one-dimensional; if $0<D<0.5$, zero-dimensional.

More specifically, in the lock-and-key mesostructure maps (\fig{maps}a,d), particles form a kinetically-trapped percolating gel structures at low temperature with the same dimensionality as the bulk. Yet, gel and bulk can be distinguished because of their different degree of order. Apart from such case, control of mesostructure geometry via the spheres radius is apparent. For small or absent spheres, growth proceeds in all directions and bulk three-dimensional mesostructures form (Fig.~S11). Increasing sphere radius hinders mesostructure growth in the medial rod direction,  leading to the formation of bilayers (Fig.~S10). When the sphere radius is larger than the radius of the rods, ribbons form because spheres cap the sides and limit growth in two directions (Fig.~S7). Finally, sufficiently large spheres frustrate growth in all direction and forms stars (Fig.~S5).

Let us remove part of the specificity of the interaction by considering symmetrically interacting rods. The corresponding mesostructure maps are shown in \fig{maps}b,e. Mesostructure formation is generally similar to the lock-and-key case discussed previously but with some important differences. Fibers now form at low temperature for larger sphere radii (Fig.~S6), since an interconnected network is no longer possible because of the too large radius of the spheres. In contrast, fibers do not form with lock-and-key rods because the rod direction vectors point in the same directions, hindering the growth of a one dimensional mesostructure in the direction of the rod axis. This behavior also limited the maximum sphere size for one-dimensional mesostructure formation in the lock-and-key case. Other differences caused by the rod interaction is the appearance of monolayers instead of bilayers, a different geometry of ribbons, which are now made of one layer only instead of two layers, and the appearance of spherical clusters instead of stars (Fig.~S4). 

Finally, we consider the case of unspecific interactions between the rods. The mesostructure maps for this case are depicted in \fig{maps}c,f. Notice that ordered mesostructures can now form at higher temperature compared to the previous cases. The number of rod beads that can interact attractively is higher, which facilitates assembly at higher temperature. Monolayers and spheres appear similarly to the symmetric case, whereas other mesostructures are unique for this simulation set. Monolayers (Fig.~S9) formed by rods with unspecific interactions can be clearly distinguished because they are the only ordered structures (\fig{maps}f). When sphere size is small and at low temperature, gel structures form (Fig.~S14). The gel structures are dynamically arrested at low temperature, but show ageing at slightly higher temperature, as can be observed by the slow decrease in potential energy (Fig.~S16). At higher temperature, these amorphous structures grow more and more compact and less connected. Eventually, they form a liquid phase (Fig.~S12). Moreover, one-dimensional worms form due to the frustration caused by presence of the spheres, but order within the fibers as found in ribbons is lost (Fig.~S8).

\subsection{Ribbon twist}
Ribbons and worms appear in our simulations when the size of the sphere is slightly larger than the radius of the rod, frustrating the formation of structures with higher dimensionality. More generally, building blocks that self-assemble one-dimensional structures have specific anisotropic geometric properties, usually a directional intermolecular potential,\cite{douglas_theoretical_2009} or they are amphiphilic and have a rigid part.\cite{hartgerink_self-assembly_2001, stupp_supramolecular_2014} The rod-sphere model of this work fulfills the latter property.  Furthermore, rod particles have been shown to self-assemble into helices due to geometrical frustration both by experiments\cite{kemp_helical_2001, srivastava_light-controlled_2010, korevaar_pathway_2012, grason_perspective_2016, hifsudheen_helix_2017} and simulations.\cite{nguyen_switchable_2009} Interestingly, we also observed signs of twisting in the ribbons self-assembled from rod-sphere particles with lock-and-key interaction.

\begin{figure*}
    \centering
	{\includegraphics[width=1\textwidth]{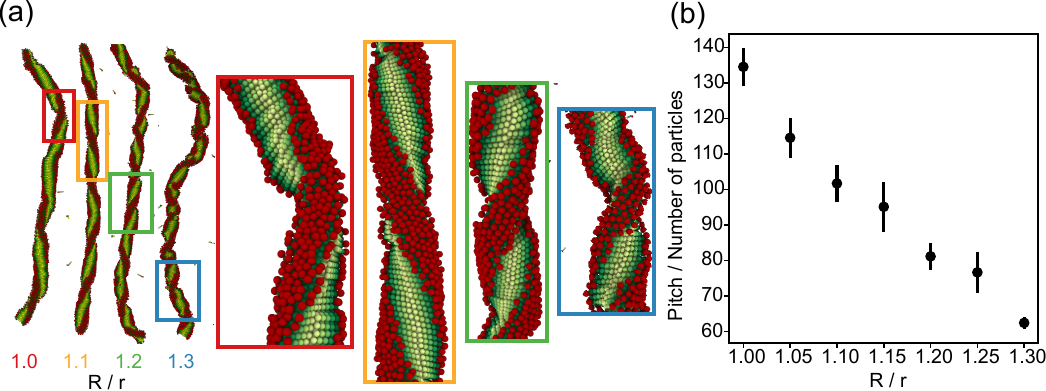}}
	\caption{\textbf{Twisted ribbons with rod-sphere particles.} (a) Snapshots of twisted ribbons for four choices of the sphere radii $R/r$. (b) The average pitch of the twisted double helix decreases as the sphere radius increases. Pitch is defined as the number of vertically-stacked particles needed to achieve one full rotation of the double helix.}
	\label{fig:twists}
\end{figure*}

To investigate the phenomenon of ribbon twisting in more detail, we initialize simulations from a straight ribbon and check for the appearance of spontaneous twisting (\fig{twists}). Simulations are performed for different values of the size ratio $R/r$ at a fixed temperature $kT = 0.5\varepsilon$. Indeed, ribbons spontaneous twist (\fig{twists}a) and form double helices. There cannot exist a  preferred handedness because particle interactions are achiral by design. Left and right handedness is found with equal probability in simulation. In addition, handedness can switch at topological defects along a single ribbon.

Double helices form because their geometry allows the attractive rods to be in contact while leaving sufficient space for the spheres to avoid overlap. For the same reason, when increasing the radius of the sphere, the pitch of the double helix decreases (\fig{twists}b). Spontaneous twisting is highly robust and even continues to $R/r=1$, in which case helix formation increases the contact number of rod beads.

\section{Conclusions}

This work investigated a coarse-grained rod-sphere model. Molecular dynamics simulations showed the self-assembly of diverse range of anisotropic mesostructures with various dimensionalities, from zero- to three-dimensional, where dimensionality was calculated with a dimensionality parameter. Since the spheres are not attractive, dimensionality is controlled solely by geometric frustration caused by the presence of spheres at the end of the rod, in particular in the case when the sphere radius is larger than the rod radius. Geometric frustration also induces the formation of double helix structures, in which the pitch is again controlled by the radius of the spheres.

This work can be of guidance for the design of nanobuilding blocks, which self-assemble into desired structures. Our results match the classical theory on surfactants and micelle formation. In addition, we shed light on the self-assembly mechanism of nanoparticles coated with \piprox or other polymers that form rigid bundles after crystallization, driven by the geometric hindrance exercised by the nanoparticles on the ligand bundles. Mesoscale structures formed by \piprox grafted silica nanoparticles, such as fibers and spherical clusters, were reproduced in our simulations. Additional predicted mesostructures at different levels of frustration and interaction parameters are not yet accessible by experiment. Moreover, while our highly coarse-grained model can reach much longer timescales than all-atom models, it does not yet reach the complete experimental assembly time, which can be in the seconds to hours \cite{nabiyan_self-assembly_2023} Finally, our rod-sphere model tethering just a single rod to a sphere is a simplification. Future extensions can include the presence of multiple rods. 

\begin{acknowledgement}
F.T., A.M., and M.E. acknowledge support by the Deutsche Forschungsgemeinschaft (DFG) Project-ID 416229255 - SFB 1411. Scientific support and HPC resources provided by the Erlangen National High Performance Computing Center (NHR@FAU) under the NHR project b168dc are gratefully acknowledged. NHR funding is provided by federal and Bavarian state authorities. NHR@FAU hardware is partially funded by DFG, Project-ID 440719683.
\end{acknowledgement}

\begin{suppinfo}
Supporting Information contains additional simulation snapshots showing all types of mesostructures.
\end{suppinfo}

\providecommand{\latin}[1]{#1}
\makeatletter
\providecommand{\doi}
  {\begingroup\let\do\@makeother\dospecials
  \catcode`\{=1 \catcode`\}=2 \doi@aux}
\providecommand{\doi@aux}[1]{\endgroup\texttt{#1}}
\makeatother
\providecommand*\mcitethebibliography{\thebibliography}
\csname @ifundefined\endcsname{endmcitethebibliography}
  {\let\endmcitethebibliography\endthebibliography}{}

\end{document}